%% file: main.tex
\documentclass[12pt]{article}
\usepackage[margin=1in]{geometry}
\usepackage[T1]{fontenc}
\usepackage{times}
\usepackage{amsmath,amssymb,amsthm}
\usepackage{graphicx}
\graphicspath{{figures/}}
\usepackage{booktabs}
\usepackage{threeparttable}
\usepackage{multirow}
\usepackage[round]{natbib}
\usepackage{hyperref}
\usepackage{url}
\newcommand{\botrule}{\bottomrule}  % compatibility with IMA table files
\numberwithin{equation}{section}

\title{Stability Anchors and Risk Amplifiers:\\ Tail Spillovers Across Stablecoin Designs}

\author{
Wenbin Wu\textsuperscript{1} \and Can Liu\textsuperscript{2} \\[12pt]
\textsuperscript{1}Cambridge Centre for Alternative Finance, Judge Business School,\\
University of Cambridge, \href{mailto:w.wu@jbs.cam.ac.uk}{w.wu@jbs.cam.ac.uk} \\[4pt]
\textsuperscript{2}Peking University HSBC Business School, \href{mailto:liucan@stu.pku.edu.cn}{liucan@stu.pku.edu.cn}
}

\date{December 2025}

\begin{document}

\maketitle

\begin{abstract}
This paper investigates systemic risk transmission across stablecoin markets using Quantile Vector Autoregression (QVAR). Analyzing eight major stablecoins with day data coverage from 2021 to 2025, supplemented by minute-level event studies on three additional coins experiencing major depegs until 2025, we document three findings. First, stabilization mechanism dictates tail-risk behavior: fiat-backed stablecoins function as ``stability anchors'' with near-zero net spillovers across quantiles, while algorithmic and crypto-collateralized designs become risk amplifiers specifically under extreme market conditions. Second, the theoretical risk isolation between fiat and crypto markets breaks down during stress: direct volatility channels emerge between the US Dollar Index and Bitcoin that bypass stablecoin intermediation. Third, Forbes-Rigobon contagion tests across four depeg events show heterogeneous transmission: after adjusting for volatility, algorithmic stablecoins exhibit significant residual contagion while fiat-backed coins show flight-to-quality effects. These findings imply that uniform stablecoin regulation is inappropriate; regulatory capital buffers for extreme losses should be 2--3$\times$ higher for non-fiat-backed stablecoins than median-based measures indicate.
\end{abstract}


\bigskip

\section{Introduction}\label{sec:intro}

Stablecoins act as the primary interface between the traditional financial system and the decentralized web, with a market capitalization exceeding \$200 billion in late 2025. While designed to function as neutral ``stability anchors'' for trading and lending, they exhibit a dual nature: acting as safe havens during mild volatility, yet potentially transforming into powerful amplifiers of systemic risk during extreme market stress. Understanding this duality is the central challenge for financial stability in the crypto ecosystem.

The systemic vulnerability of stablecoins is structurally diverse. Algorithmic and crypto-collateralized stablecoins, such as the failed BitUSD \citep{moin2020sok,bitmex2018stablecoins} or TerraUSD (UST), rely on endogenous arbitrage mechanisms and volatile collateral. When confidence evaporates, these mechanisms can trigger ``death spirals,'' as seen in the \$40 billion collapse of UST in 2022 \citep{liu2023anatomy,briola2023anatomy}. Conversely, fiat-backed stablecoins like USD Coin (USDC) face exogenous risks from the traditional banking sector, evidenced by the 2023 depeg following the Silicon Valley Bank failure \citep{oefele2024flight}. These events are not merely historical anomalies; they represent distinct failure modes where risk transmission channels shift abruptly under stress \citep{klages-mundt2020instability}.

Current literature has extensively documented the drivers of stablecoin instability, focusing on reserve adequacy and leverage \citep{gorton2023leverage,gadzinski2024break}. \citet{yip2022event} applies Diebold-Yilmaz spillover analysis to the May 2022 UST crash, documenting heterogeneous contagion patterns across fiat-backed, crypto-collateralized, and algorithmic stablecoins. However, their analysis is limited to a single crisis episode and relies on mean-based VAR methods that fail to capture the asymmetry of risk transmission across market conditions. We lack a systematic understanding of which specific designs function as ``risk sinks'' versus ``risk sources'' when the market moves to extreme quantiles.

To address this gap, this paper employs a Quantile Vector Autoregression (QVAR) framework to map the changing topology of stablecoin risk networks. We contribute to the literature in three ways:

\textit{First, we quantify how spillover amplification varies across the return distribution.} While prior work using mean-based methods documents that mechanism type affects spillover patterns \citep{yip2022event}, we provide the first QVAR-based estimates at the 5th, 50th, and 95th percentiles, revealing that fiat-backed stablecoins maintain near-zero net spillover indices across all quantiles, while algorithmic and crypto-collateralized designs show 15--50 percentage point amplification specifically in the tails. 

\textit{Second, we construct a three-tier fiat-stablecoin-crypto network to map cross-market transmission.} While the inverse relationship between the US Dollar Index and Bitcoin is documented in prior work, we extend this by integrating both into a unified QVAR framework with stablecoins. This reveals that the theoretical barrier between traditional and crypto markets, established by \citet{liu2021risks}, collapses during extreme conditions, with direct volatility channels emerging that bypass stablecoin intermediation entirely. Spillover magnitudes increase by up to 40 percentage points in tails relative to normal conditions.

\textit{Third, we apply Forbes-Rigobon contagion tests to distinguish true contagion from interdependence.} While prior studies use BEKK or DCC-GARCH models on stablecoin depegs, we employ Forbes-Rigobon adjusted correlations, which explicitly correct for heteroskedasticity bias, across four major events: UST collapse (May 2022), SVB/USDC crisis (March 2023), sUSD depeg (April 2025), and USDe crash (October 2025). After controlling for volatility increases, algorithmic stablecoins show significant residual contagion while fiat-backed coins exhibit flight-to-quality dynamics.

The remainder of this paper is organized as follows. Section~\ref{sec:literature} reviews the related literature. Section~\ref{sec:data} describes our data. Section~\ref{sec:methodology} presents the QVAR methodology. Section~\ref{sec:results} reports our empirical results, including rolling window analysis, three-tier network analysis, high-frequency event studies, and robustness checks. Section~\ref{sec:conclusion} concludes with policy implications.

\section{Literature Review}\label{sec:literature}

Research on stablecoin risk has evolved from analyzing isolated design mechanisms to mapping complex cross-market contagions. This section synthesizes three streams of literature: mechanism-specific vulnerabilities, systemic spillover effects, and methodological advances in tail-risk measurement.

Regarding mechanism design and vulnerability, early studies established that risk exposure is fundamentally dictated by collateral composition \citep{adachi2022stablecoins,macdonald2022stablecoins}. \citet{gorton2023leverage} and \citet{gadzinski2024break} demonstrate that while fiat-backed stablecoins function like narrow banks, vulnerable to traditional runs \citep{bertsch2023stablecoins,anadu2024runs}, algorithmic variants face distinct, endogenous fragility driven by speculative attacks and loss of confidence. The collapse of UST validated theoretical models by \citet{klages-mundt2020instability}, proving that arbitrage mechanisms relying on volatile endogenous collateral enter ``death spirals'' once a critical depegging threshold is breached \citep{liu2023anatomy,yip2022event}. Distinct price behaviors between these types during turbulence confirm this structural heterogeneity \citep{deblasis2023intelligent}.

Beyond idiosyncratic failure, stablecoins act as vectors for systemic contagion. \citet{briola2023anatomy} and \citet{lee2023dissecting} show that major depeg events trigger immediate volatility spillovers across the broader crypto ecosystem. Crucially, this transmission is bidirectional: \citet{wu2023asset} and \citet{kim2022dk} find that large-scale reserve adjustments by stablecoin issuers can transmit shockwaves back to traditional short-term credit markets. Risk transmission occurs through multiple channels, including arbitrage, collateral liquidation, and on-chain bridging \citep{badev2023interconnected,sween2023contagion}. During financial turmoil, stablecoins frequently serve as ``flight-to-safety'' destinations, yet paradoxically become contagion hubs when the safety itself is questioned \citep{gubareva2023xn,vukovic2024spillovers,huang2023jk,iyer2023new}.

On the methodological front, traditional mean-based models (VAR, GARCH) fail to capture the asymmetric nature of these spillovers. As summarized in Table~\ref{tab:spillover_methods}, the field is shifting toward quantile-based approaches to isolate tail dependencies. While early work focused on average-level co-movements \citep{giudici2020ke,let2023gy}, recent applications of quantile Granger causality and Cross-Quantilograms reveal that risk transmission structures change fundamentally under extreme conditions \citep{fernandez-mejia2024ig,kolodziejczyk2023hc,paeng2024xk}. Furthermore, limitations in standardizing on-chain data and bridging ratios have historically hampered precise risk quantification \citep{watsky2024primary,aldasoro2024stablecoins}, a gap this study addresses through robust quantile estimation.

\input{tables/spillover_methods_table.tex}

This paper advances this methodological trajectory. Unlike prior studies that treat stablecoins as a homogeneous asset class or rely on normal distribution assumptions, we employ Quantile VAR to explicitly distinguish between \textit{risk absorption} and \textit{risk amplification} roles across the full distribution of market states.

\section{Data Description and Preliminary Analysis}\label{sec:data}

\subsection{Data Sources and Processing}

We collect data on 11 major stablecoins spanning three mechanism types: fiat-backed, crypto-collateralized, and algorithmic. The full sample period spans December 2020 to November 2025, comprising 1,837 trading days. Price data are sourced from CoinMetrics Pro using the ReferenceRateUSD metric, which provides volume-weighted aggregated prices across major exchanges. We study the deviation of stablecoin prices from the \$1 peg, using basis point differenced series to reflect daily changes.

For the QVAR analysis requiring balanced panels, we focus on eight stablecoins with sufficient overlapping data: USDT, USDC, BUSD, TUSD, DAI, LUSD, FRAX, and sUSD. Three coins are excluded from the main QVAR estimation due to data constraints: UST (collapsed May 2022, only 341 overlapping observations), FDUSD (launched June 2023), and USDe (launched February 2024). However, these excluded coins, which experienced some of the most dramatic depeg events, are analyzed separately through high-frequency event studies in Section~\ref{sec:event_study}.

\begin{figure}[!t]
\centering
\includegraphics[width=\textwidth]{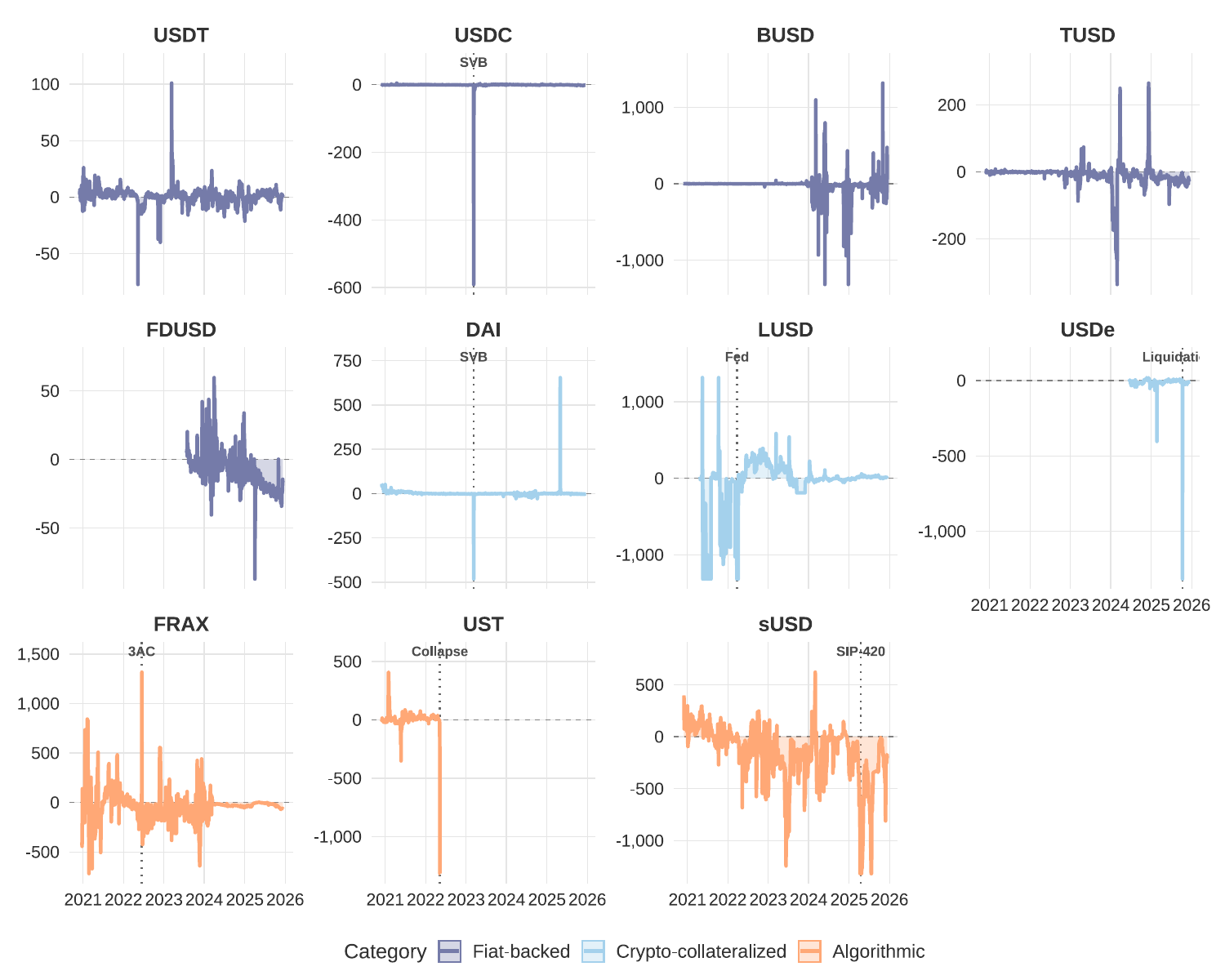}
\caption{Time series characteristics of major stablecoin deviation value changes. Vertical dotted lines indicate major depeg events annotated by cause; see Table~\ref{tab:depeg_events} for details.}\label{fig:stablecoin_diff}
\end{figure}

Figure~\ref{fig:stablecoin_diff} displays deviation time series for all 11 stablecoins. Volatility increases systematically from fiat-backed to crypto-collateralized to algorithmic types. Notable extreme events include the UST collapse, USDe crash to \$0.68 , and sUSD depeg. The pronounced volatility clustering and fat tails motivate our quantile-based approach.

\input{tables/depeg_events_table.tex}

Table~\ref{tab:depeg_events} categorizes depeg triggers into three types: \textit{Macro} shocks originate from external events in traditional finance or broader crypto markets (e.g., SVB collapse, Fed tightening); \textit{Contagion} events reflect spillovers from other stablecoins (e.g., DAI's exposure to USDC via the Peg Stability Module); \textit{Endogenous} failures stem from internal protocol design flaws (e.g., UST's death spiral, sUSD's broken arbitrage incentives). This taxonomy informs our subsequent analysis of mechanism-specific spillover patterns.

Table~\ref{tab:deviation_stats} reports summary statistics. USDT exhibits the tightest peg (std dev: 7.01 bp), while algorithmic coins show order-of-magnitude higher volatility (sUSD: 286 bp). The extreme minima for USDe ($-3197$ bp) and UST ($-3768$ bp) highlight tail risk in non-fiat-backed designs.

\begin{table}[!t]
\caption{Summary statistics of stablecoin price deviations (basis points)\label{tab:deviation_stats}}
\begin{tabular*}{\columnwidth}{@{\extracolsep\fill}lrrrrrrr@{\extracolsep\fill}}
\toprule
Stablecoin & Mkt Cap (\$B) & Mean & Std Dev & Min & Median & Max \\
\midrule
\multicolumn{7}{l}{\textit{Fiat-backed (Total: \$269.2B)}} \\
USDT & 188.7 & 1.01 & 7.01 & $-77.36$ & 1.28 & 100.98 \\
USDC & 76.4 & $-0.81$ & 15.71 & $-590.86$ & $-0.23$ & 4.91 \\
BUSD & 0.06 & $-20.48$ & 139.58 & $-1829.31$ & $-0.37$ & 3387.70 \\
TUSD & 0.49 & $-10.37$ & 33.80 & $-336.00$ & $-3.04$ & 265.04 \\
FDUSD & 3.5 & $-8.73$ & 13.24 & $-87.21$ & $-8.45$ & 59.73 \\
\midrule
\multicolumn{7}{l}{\textit{Crypto-collateralized (Total: \$11.2B)}} \\
DAI & 4.4 & 0.90 & 25.09 & $-482.24$ & $-0.47$ & 654.39 \\
LUSD & 0.28 & $-84.34$ & 486.77 & $-3593.81$ & 1.09 & 5543.79 \\
USDe & 6.6 & $-14.52$ & 138.36 & $-3196.76$ & $-8.49$ & 19.84 \\
\midrule
\multicolumn{7}{l}{\textit{Algorithmic (Total: \$0.7B)}} \\
FRAX & 0.68 & $-32.95$ & 142.22 & $-718.86$ & $-29.98$ & 1523.67 \\
UST & \textemdash & $-3.95$ & 222.08 & $-3768.28$ & 6.40 & 408.58 \\
sUSD & 0.05 & $-156.49$ & 286.00 & $-2257.66$ & $-88.39$ & 620.77 \\
\bottomrule
\end{tabular*}
\begin{tablenotes}
\item Note: Market caps as of November 2025; UST collapsed in May 2022 and is no longer functional. Deviations in basis points: $(P_{i,t} - 1) \times 10000$. Sample: December 2020--November 2025.
\end{tablenotes}
\end{table}

\begin{figure}[!t]
\centering
\includegraphics[width=\textwidth]{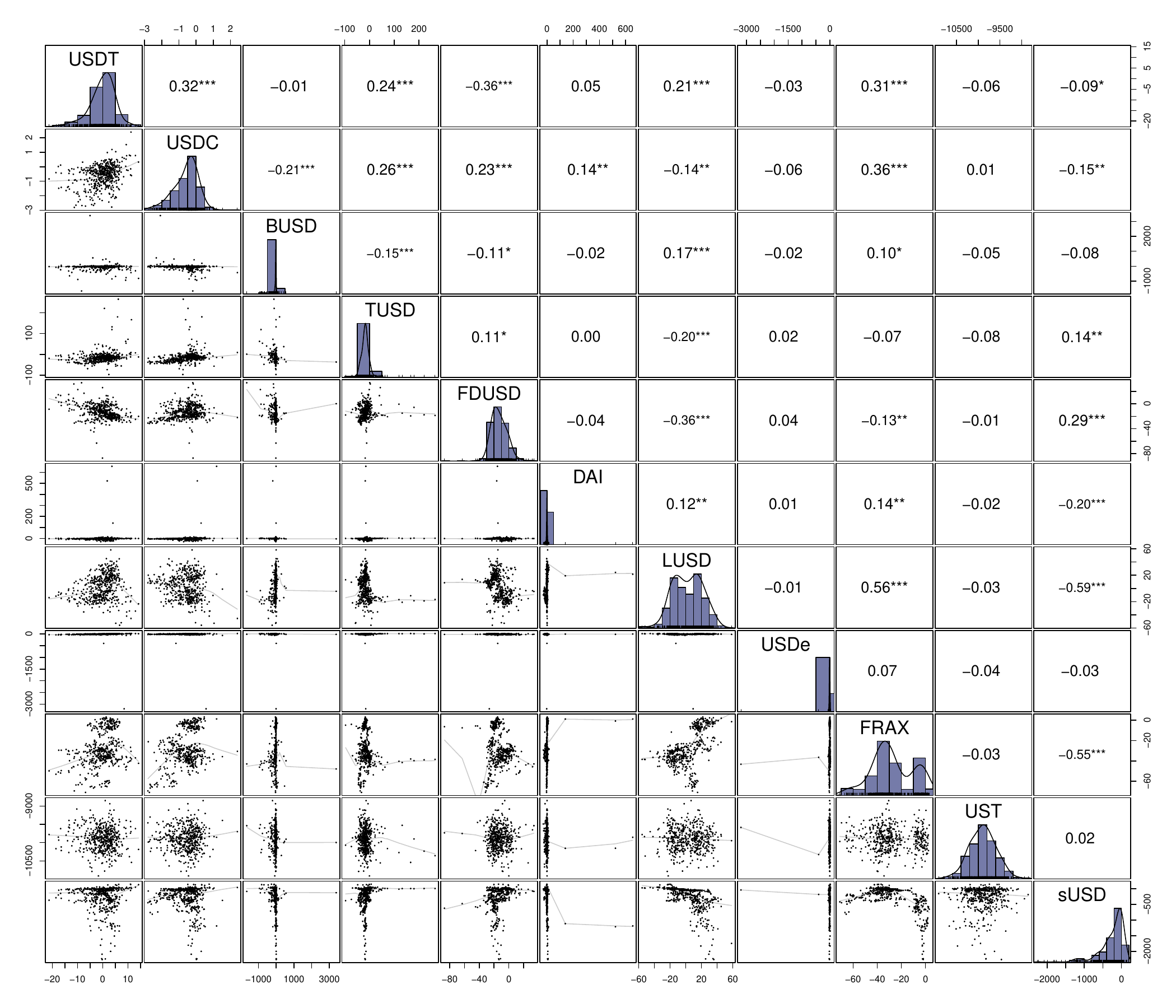}
\caption{Correlation matrix analysis of major stablecoin deviation value changes}\label{fig:stablecoin_corr}
\end{figure}

Figure~\ref{fig:stablecoin_corr} displays the correlation matrix of deviation changes. Correlations are generally low under normal conditions, but exhibit clear stratification by mechanism: decentralized stablecoins such as LUSD-FRAX show stronger co-movement than fiat-backed pairs. 

\section{Methodology}\label{sec:methodology}

\begin{figure}[!t]
\centering
\includegraphics[width=\textwidth]{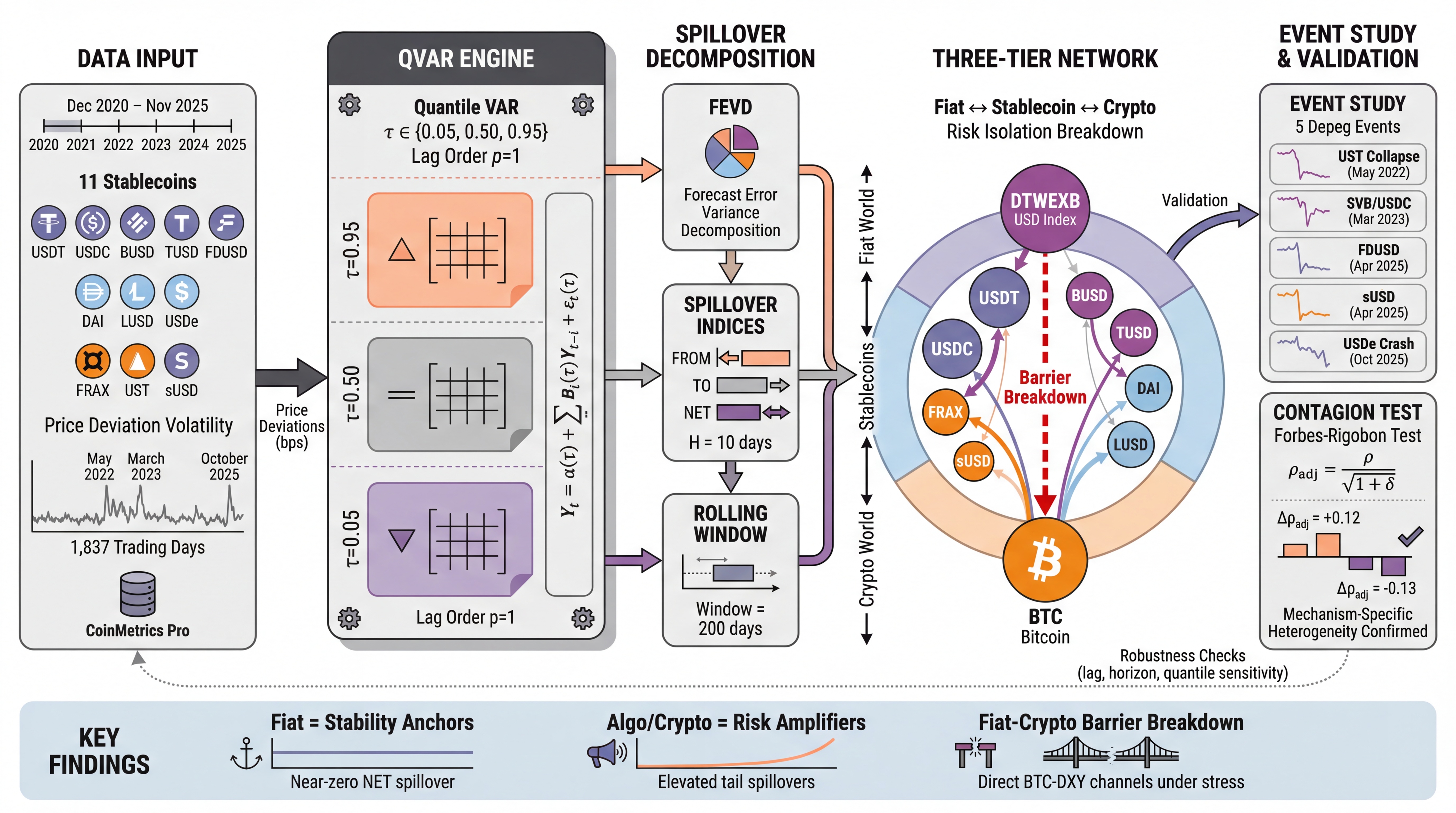}
\caption{Methodological framework overview. The analysis pipeline flows from data input through the QVAR model estimated at three quantiles, to spillover decomposition.}\label{fig:methodology_overview}
\end{figure}

Figure~\ref{fig:methodology_overview} provides an overview of our analytical framework. The pipeline flows from data input through QVAR modeling at three quantiles to spillover decomposition and network construction.

\subsection{Quantile Vector Autoregression (QVAR) Model}

We employ the Quantile Vector Autoregression (QVAR) method to measure spillover effects in stablecoin markets, capturing nonlinear characteristics of risk transmission under different market conditions. The QVAR method extends traditional vector autoregression models by allowing estimation of system dynamics at different quantile levels, revealing tail risk transmission mechanisms that traditional mean regression methods cannot capture.

The QVAR model provides a more comprehensive understanding of risk transmission in stablecoin markets by estimating system dynamics at specific quantile levels. This method was developed by \citet{white2015qvar} and applied to financial systemic risk research by \citet{chavleishvili2023rp}. For a vector $Y_t$ containing $n$ stablecoin returns, the QVAR(p) model at quantile $\tau$ is expressed as:

\begin{equation}
Y_t = \alpha(\tau) + \sum_{i=1}^p B_i(\tau) Y_{t-i} + \varepsilon_t(\tau)
\end{equation}

where $\alpha(\tau)$ is an $n \times 1$ intercept vector, $B_i(\tau)$ is an $n \times n$ coefficient matrix, $\varepsilon_t(\tau)$ is an $n \times 1$ error vector satisfying $Q_{\tau}(\varepsilon_t(\tau)|Y_{t-1}, Y_{t-2}, \ldots) = 0$, meaning the $\tau$ quantile of the error term is zero conditional on past information. $\tau \in (0,1)$ represents the quantile level of interest.

This study focuses on three key quantiles: $\tau = 0.05$ (left tail, representing spillover effects during extreme market downturns), $\tau = 0.5$ (median, representing spillover effects under normal market conditions), and $\tau = 0.95$ (right tail, representing spillover effects during extreme market upturns).

The QVAR model is estimated through quantile regression, minimizing weighted absolute deviations:

\begin{equation}
\hat{B}_1(\tau), \ldots, \hat{B}_p(\tau) = \arg\min_{B_1, \ldots, B_p} \sum_{t=p+1}^T \sum_{j=1}^n \rho_{\tau} \left( Y_{j,t} - \alpha_j(\tau) - \sum_{i=1}^p e_j' B_i(\tau) Y_{t-i} \right)
\end{equation}

where $\rho_{\tau}(u) = u(\tau - I(u < 0))$ is the quantile loss function, $I(\cdot)$ is the indicator function, and $e_j$ is the $j$-th unit vector.

\subsection{Forecast Error Variance Decomposition under QVAR}

Once QVAR model parameters are estimated, we quantify spillover effects among stablecoins by computing Forecast Error Variance Decomposition (FEVD). First, we convert the QVAR model to a Quantile Vector Moving Average (QVMA) representation:

\begin{equation}
Y_t = \mu(\tau) + \sum_{j=0}^{\infty} A_j(\tau) \varepsilon_{t-j}(\tau)
\end{equation}

where $A_0(\tau) = I_n$ (the $n$-order identity matrix), $\mu(\tau)$ is a constant vector, and $A_j(\tau)$ is the QVMA coefficient matrix computed recursively:

\begin{equation}
A_j(\tau) = \sum_{i=1}^{\min(j,p)} B_i(\tau) A_{j-i}(\tau), \quad j \geq 1
\end{equation}

For forecast horizon $H$, the contribution of stablecoin $k$ to the forecast error variance of stablecoin $j$, $\Theta_{j,k}^H(\tau)$, is calculated as:

\begin{equation}
\Theta_{j,k}^H(\tau) = \frac{\sum_{h=0}^{H-1} \left( e_j' A_h(\tau) \Sigma(\tau) e_k \right)^2 / \sigma_{kk}(\tau)}{\sum_{h=0}^{H-1} e_j' A_h(\tau) \Sigma(\tau) A_h(\tau)' e_j}
\end{equation}

where $\Sigma(\tau)$ is the error term covariance matrix at quantile $\tau$, $\sigma_{kk}(\tau)$ is the $k$-th diagonal element of $\Sigma(\tau)$, and $e_j$ and $e_k$ are the $j$-th and $k$-th unit vectors respectively.

To ensure FEVD matrix row sums equal 1, we normalize $\Theta_{j,k}^H(\tau)$:

\begin{equation}
\widetilde{\Theta}_{j,k}^H(\tau) = \frac{\Theta_{j,k}^H(\tau)}{\sum_{k=1}^n \Theta_{j,k}^H(\tau)}
\end{equation}

The normalized FEVD matrix $\widetilde{\Theta}_{j,k}^H(\tau)$ represents the proportion of stablecoin $j$'s forecast error variance contributed by shocks from stablecoin $k$ at quantile $\tau$.

\subsection{Quantile Spillover Index Construction}

Based on the FEVD matrix, we construct three key directional spillover indicators. For quantile level $\tau$:

Total inflow spillover received by stablecoin $j$ from other stablecoins (FROM):
\begin{equation}
SO^{IN}_{j}(\tau) = \sum_{k=1,k\neq j}^n \widetilde{\Theta}_{j,k}^H(\tau)
\end{equation}

Total outflow spillover from stablecoin $j$ to other stablecoins (TO):
\begin{equation}
SO^{OUT}_{j}(\tau) = \sum_{k=1,k\neq j}^n \widetilde{\Theta}_{k,j}^H(\tau)
\end{equation}

Net spillover of stablecoin $j$ (NET):
\begin{equation}
NSO_{j}(\tau) = SO^{OUT}_{j}(\tau) - SO^{IN}_{j}(\tau)
\end{equation}

The system total spillover index is calculated as the sum of off-diagonal elements divided by the sum of all elements:
\begin{equation}
SO^{Total}(\tau) = \frac{\sum_{j=1}^n \sum_{k=1,k\neq j}^n \widetilde{\Theta}_{j,k}^H(\tau)}{\sum_{j=1}^n \sum_{k=1}^n \widetilde{\Theta}_{j,k}^H(\tau)} = \frac{\sum_{j=1}^n \sum_{k=1,k\neq j}^n \widetilde{\Theta}_{j,k}^H(\tau)}{n}
\end{equation}

This index ranges from $[0,1]$ or $[0\%,100\%]$ when expressed as a percentage, with higher values indicating stronger mutual influence among stablecoins in the system.

\subsection{Quantile Relative Spillover Index}

To capture asymmetry in spillover effects under different market conditions, we define quantile relative spillover indices measuring additional spillover effects of tail risk relative to median conditions. The left-tail relative spillover index is:
\begin{equation}
\Delta SO^{L}_{j} = SO_{j}(0.05) - SO_{j}(0.5)
\end{equation}

The right-tail relative spillover index is:
\begin{equation}
\Delta SO^{R}_{j} = SO_{j}(0.95) - SO_{j}(0.5)
\end{equation}

where $SO_{j}(\tau)$ can represent $SO^{IN}_{j}(\tau)$, $SO^{OUT}_{j}(\tau)$, or $NSO_{j}(\tau)$, corresponding to inflow, outflow, or net spillover respectively. Positive relative spillover values indicate strengthened spillover effects under tail risk conditions, while negative values indicate weakening.

This quantile relative spillover analysis method can identify which stablecoins become major risk sources or receivers under extreme market conditions, which is crucial for regulators and market participants to understand systemic risk transmission mechanisms. \citet{adrian2019oq} emphasized the importance of tail risk when studying financial vulnerability, and our quantile framework aligns with this approach.

\subsection{Rolling Window Analysis}

To capture time-varying characteristics of spillover effects, we employ rolling window methods for QVAR analysis. Specifically, for total sample length $T$, we use fixed window length $w$ and step size $s$:

\begin{equation}
\text{Window}_i = \{t: i \cdot s + 1 \leq t \leq i \cdot s + w\}, \quad i = 0, 1, 2, \ldots, \lfloor(T-w)/s\rfloor
\end{equation}

For each window, we estimate QVAR models at three quantiles ($\tau \in \{0.05, 0.5, 0.95\}$) and calculate corresponding spillover indices, obtaining dynamic paths of spillover effects over time. This method captures structural changes in stablecoin market risk transmission patterns, particularly during market stress events.

\section{Empirical Results}\label{sec:results}

Our empirical analysis establishes three findings. First, stabilization mechanism, not market size, determines whether a stablecoin absorbs or amplifies systemic risk. Second, the theoretical isolation between fiat and crypto markets breaks down precisely when it matters most: during tail events. Third, this mechanism-specific heterogeneity reflects genuine causal differences, not merely correlation patterns. We present evidence for each finding in turn.

\subsection{Finding 1: Mechanism Determines Spillover Behavior}

The mechanism-spillover relationship has economic foundations. Fiat-backed stablecoins absorb shocks through reserve buffers: a \$100M redemption draws on \$100M in T-bills or bank deposits, not market sales. Algorithmic and crypto-collateralized designs lack this buffer: maintaining the peg requires market transactions (minting/burning governance tokens, liquidating collateral), which transmit rather than absorb volatility. This predicts that non-fiat-backed coins should show elevated \textit{outflow} spillovers specifically during stress, when arbitrage mechanisms are most active.

\begin{figure}[!t]
\centering
\includegraphics[width=0.95\textwidth]{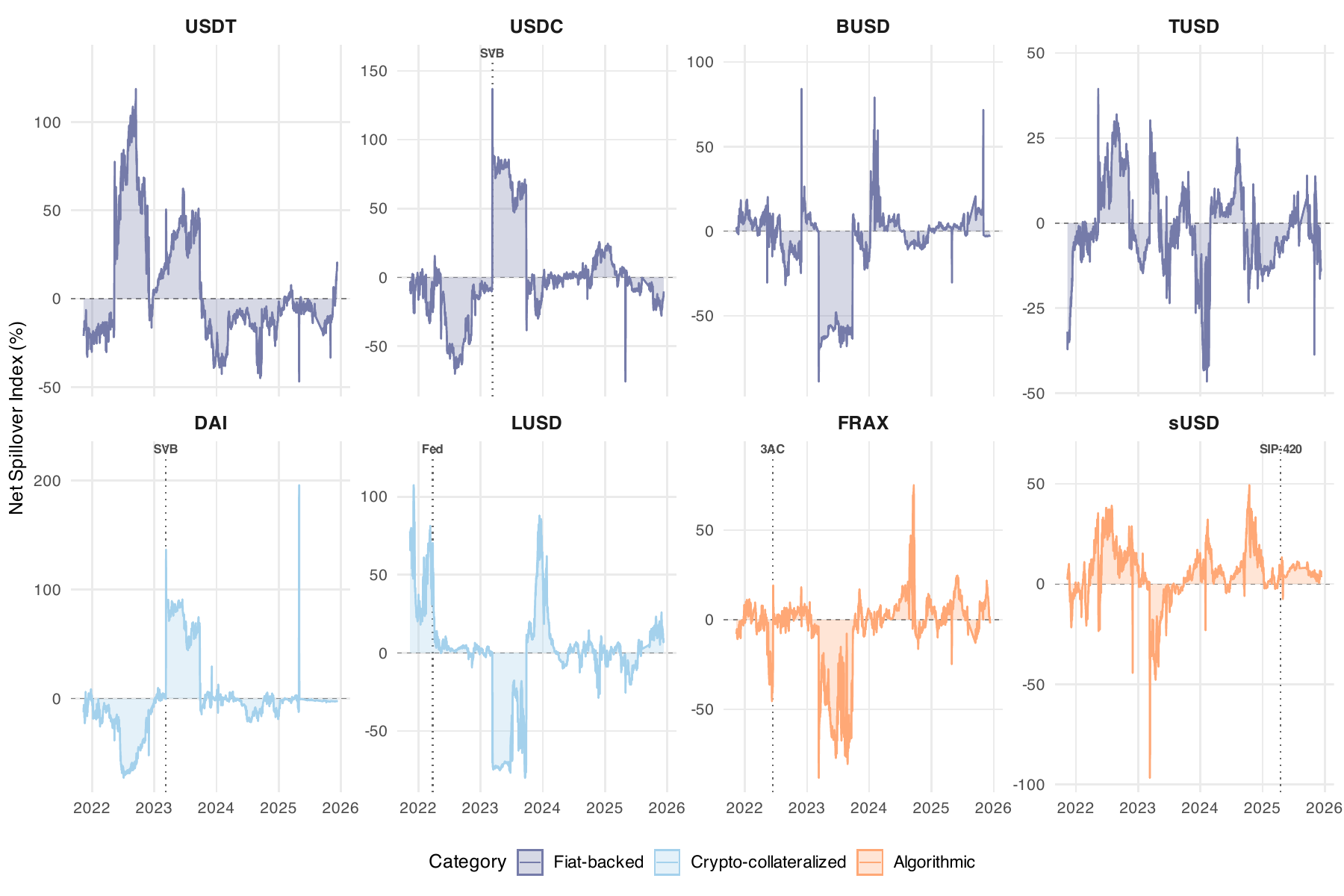}
\caption{Dynamic paths of rolling window net spillover indices for major stablecoins}\label{fig:rolling_net_spillover}
\end{figure}

We begin by documenting that fiat-backed stablecoins function as ``stability anchors'' while algorithmic and crypto-collateralized designs become risk amplifiers under stress. Figure~\ref{fig:rolling_net_spillover} displays rolling window net spillover indices, revealing a stark divide: fiat-backed stablecoins (USDT, USDC, BUSD, TUSD) maintain near-zero net spillovers throughout the sample, while algorithmic coins (FRAX, sUSD) and crypto-collateralized coins (DAI, LUSD) exhibit dramatic volatility, alternating between major risk transmitters and receivers during stress periods.

This pattern intensifies during market stress: system-wide spillovers spike during three crisis periods---H1 2022 (UST collapse), Q2-Q3 2023 (SVB crisis), and Q1 2025 (crypto rally volatility)---with tail quantile spillovers consistently exceeding median spillovers by 15--30 percentage points, consistent with \citet{fernandez-mejia2024ig}.

\begin{figure}[!t]
\centering
\includegraphics[width=0.95\textwidth]{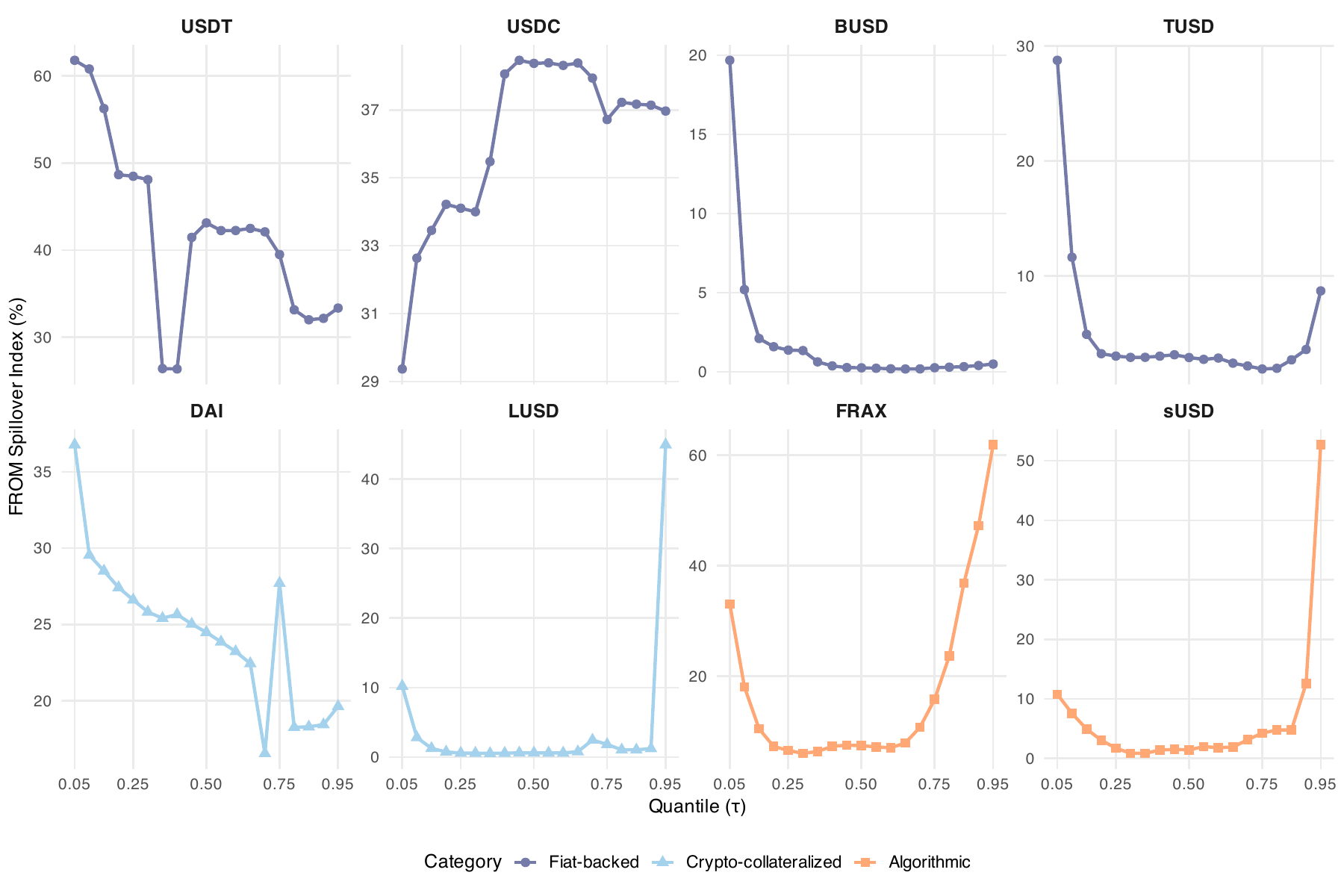}
\caption{Quantile inflow (FROM) index curves for each stablecoin}\label{fig:from_spillover_curve}
\end{figure}

\begin{figure}[!t]
\centering
\includegraphics[width=0.95\textwidth]{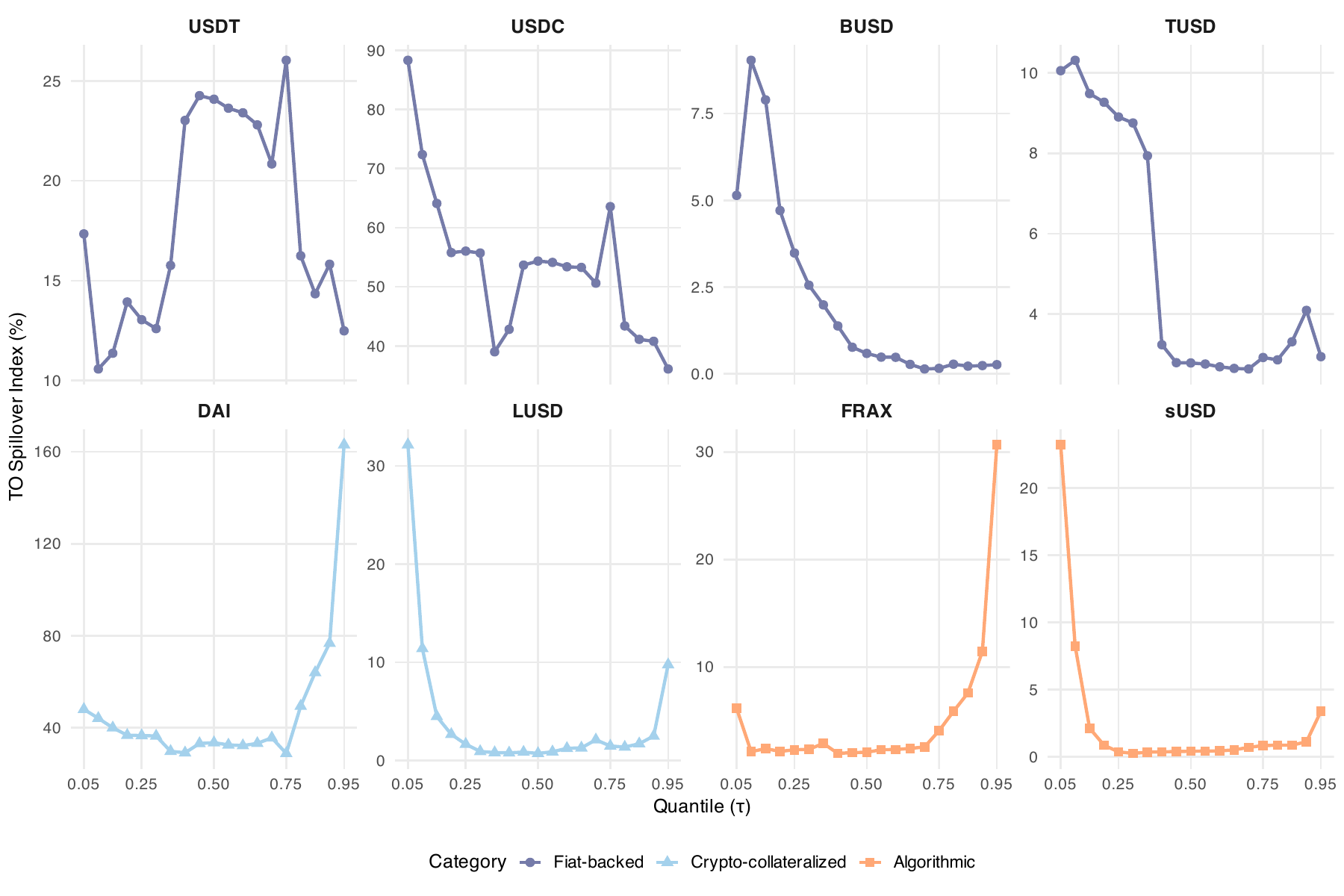}
\caption{Quantile outflow (TO) index curves for each stablecoin}\label{fig:to_spillover_curve}
\end{figure}

Figures~\ref{fig:from_spillover_curve} and \ref{fig:to_spillover_curve} decompose spillovers into inflow (FROM) and outflow (TO) components across quantiles. The key insight is structural: cross-coin variation in spillover levels far exceeds within-coin variation across quantiles. Fiat-backed stablecoins maintain low outflow indices regardless of market conditions, while algorithmic and crypto-collateralized coins show pronounced tail sensitivity, becoming both major risk transmitters and receivers during stress. This asymmetry is consistent with \citet{deblasis2023intelligent}'s finding that mechanism design shapes crisis behavior.

\begin{figure}[!t]
\centering
\includegraphics[width=0.98\textwidth]{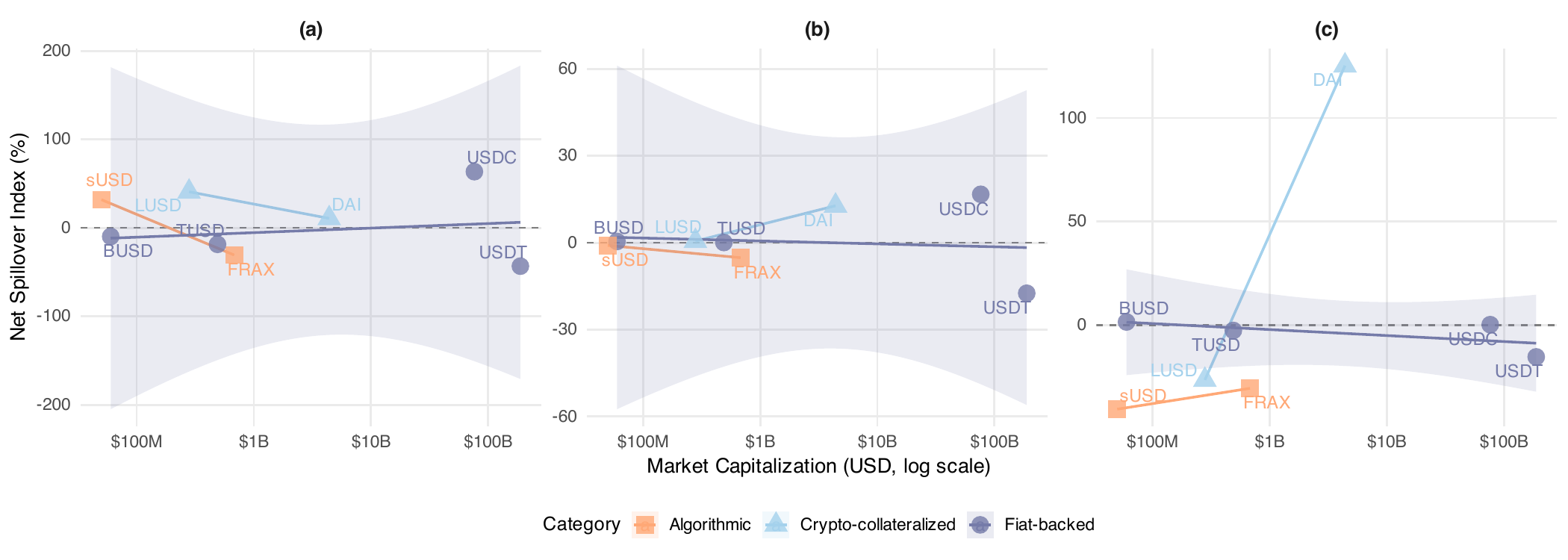}
\caption{Relationship between market capitalization and net spillover effects under different quantiles. (a) $\tau=0.05$, (b) $\tau=0.5$, (c) $\tau=0.95$. Shading represents confidence intervals; different shapes represent stablecoin types.}\label{fig:qvar_mcap_spillover}
\end{figure}

A natural alternative hypothesis is that market capitalization, not mechanism, drives these patterns---larger coins might simply be more stable. Figure~\ref{fig:qvar_mcap_spillover} rejects this explanation. The flat regression lines show no relationship between market cap and net spillover; instead, coins cluster by mechanism type. USDT (\$189B) and USDC (\$76B) behave identically to smaller fiat-backed coins, while tiny algorithmic coins (sUSD, \$50M) show the same amplification patterns as larger crypto-collateralized coins (DAI, \$4.4B).

\subsection{Finding 2: Fiat-Crypto Isolation Breaks Down Under Stress}

The previous analysis treats stablecoins as a closed system. But stablecoins exist precisely to bridge traditional finance and crypto markets---does this bridge function create contagion channels during stress? To answer this, we extend our QVAR network to include Bitcoin (representing crypto) and the US Dollar Index (representing fiat), constructing a three-tier ``fiat-stablecoin-crypto'' network.

\begin{figure}[!t]
\centering
\includegraphics[width=0.98\textwidth]{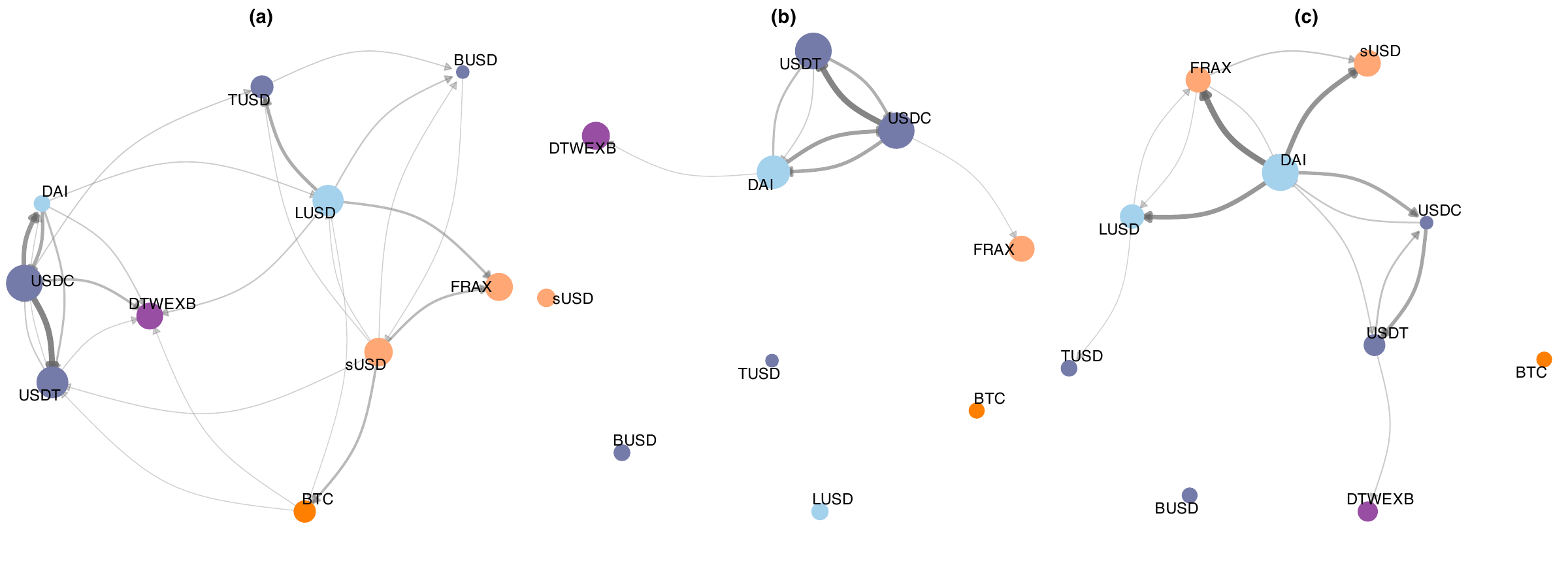}
\caption{Stablecoin spillover network structure under extreme quantiles. (a) $\tau=0.05$, (b) $\tau=0.5$, (c) $\tau=0.95$. Nodes represent different stablecoins; edge direction and thickness indicate spillover effect direction and strength.}\label{fig:extended_spillover_network}
\end{figure}

Figure~\ref{fig:extended_spillover_network} reveals a striking pattern. Under normal conditions ($\tau=0.5$), the network exhibits effective compartmentalization: fiat-stablecoin connections are weak, and the US Dollar Index sits at the periphery with minimal spillover links. Under extreme conditions ($\tau=0.05$ or $0.95$), this isolation collapses. The US Dollar Index becomes embedded in the contagion network, with direct spillover channels to Bitcoin that bypass stablecoin intermediation entirely. This barrier breakdown represents a novel systemic risk channel, extending \citet{briola2023anatomy}'s findings on crypto-stablecoin linkages to the broader fiat-crypto interface.

\begin{figure}[!t]
\centering
\includegraphics[width=0.98\textwidth]{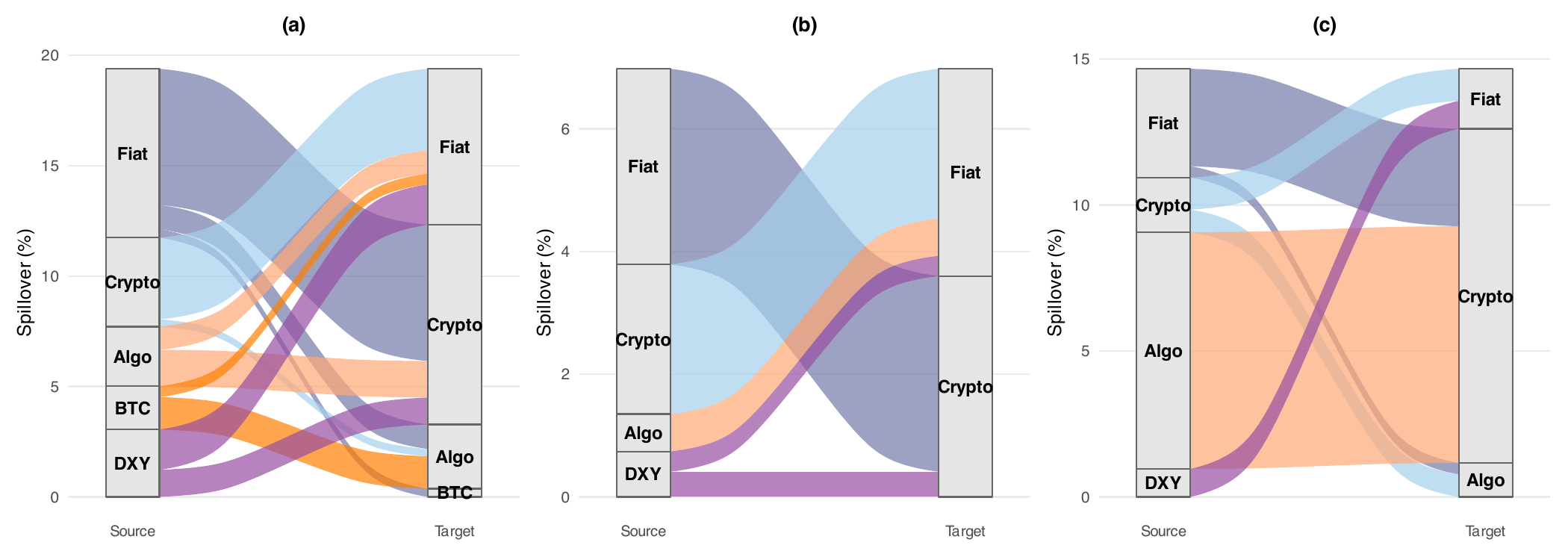}
\caption{Inter-category spillover flows under different quantiles. (a) $\tau=0.05$ (left tail), (b) $\tau=0.50$ (median), (c) $\tau=0.95$ (right tail). Flow width represents spillover magnitude between asset categories.}\label{fig:spillover_sankey}
\end{figure}

Figure~\ref{fig:spillover_sankey} quantifies these cross-category flows. Under normal conditions, spillovers are minimal and predominantly intra-category. Under tail conditions, cross-category transmission intensifies dramatically: algorithmic and crypto-collateralized stablecoins serve as conduits channeling risk between the fiat (DXY) and crypto (BTC) anchors, while fiat-backed stablecoins remain relatively insulated.

Table~\ref{tab:top_deltas} identifies the specific transmission channels that activate under stress. The top spillover changes involve crypto-collateralized and algorithmic stablecoins (LUSD, DAI, sUSD, FRAX), with spillover increases of 20--40 percentage points relative to normal conditions. Crucially, the US Dollar Index (DTWEXB) appears in several top pairs---USDC$\rightarrow$DTWEXB, LUSD$\rightarrow$DTWEXB, USDT$\rightarrow$DTWEXB---confirming that tail events activate direct fiat-crypto transmission channels.

\begin{table}[!t]
\caption{Top ten spillover effect differences between extreme quantiles and median\label{tab:top_deltas}}
\begin{tabular*}{\columnwidth}{@{\extracolsep\fill}lclc@{\extracolsep\fill}}
\toprule
Spillover Direction & $\Delta SI^L$ & Spillover Direction & $\Delta SI^R$ \\
\midrule
LUSD $\rightarrow$ TUSD & 19.96 & DAI $\rightarrow$ FRAX & 41.25 \\
LUSD $\rightarrow$ FRAX & 13.58 & DAI $\rightarrow$ sUSD & 33.73 \\
sUSD $\rightarrow$ FRAX & 13.57 & DAI $\rightarrow$ LUSD & 32.34 \\
sUSD $\rightarrow$ BTC$^\dagger$ & 13.47 & FRAX $\rightarrow$ sUSD & 7.37 \\
USDC $\rightarrow$ DTWEXB$^\dagger$ & 13.18 & FRAX $\rightarrow$ DAI & 4.07 \\
USDC $\rightarrow$ DAI & 11.63 & USDT $\rightarrow$ DTWEXB$^\dagger$ & 3.81 \\
USDC $\rightarrow$ USDT & 8.78 & LUSD $\rightarrow$ FRAX & 3.07 \\
LUSD $\rightarrow$ BUSD & 7.06 & FRAX $\rightarrow$ LUSD & 2.57 \\
LUSD $\rightarrow$ DTWEXB$^\dagger$ & 6.36 & LUSD $\rightarrow$ TUSD & 2.39 \\
DAI $\rightarrow$ LUSD & 4.59 & DAI $\rightarrow$ USDC & 2.21 \\
\bottomrule
\end{tabular*}
\begin{tablenotes}
\item Note: $^\dagger$ indicates Bitcoin or US Dollar Index.
\end{tablenotes}
\end{table}

\subsection{Finding 3: Causal Evidence from Crisis Events}\label{sec:event_study}

The preceding analysis documents correlation patterns. But do these reflect causal spillover channels, or merely common factor exposures? We address this through high-frequency event studies around four major depeg events, using minute-level data and econometric techniques designed to distinguish true contagion from volatility-induced correlation increases.

\begin{figure}[!t]
\centering
\begin{tabular}{cc}
\includegraphics[width=0.34\textwidth]{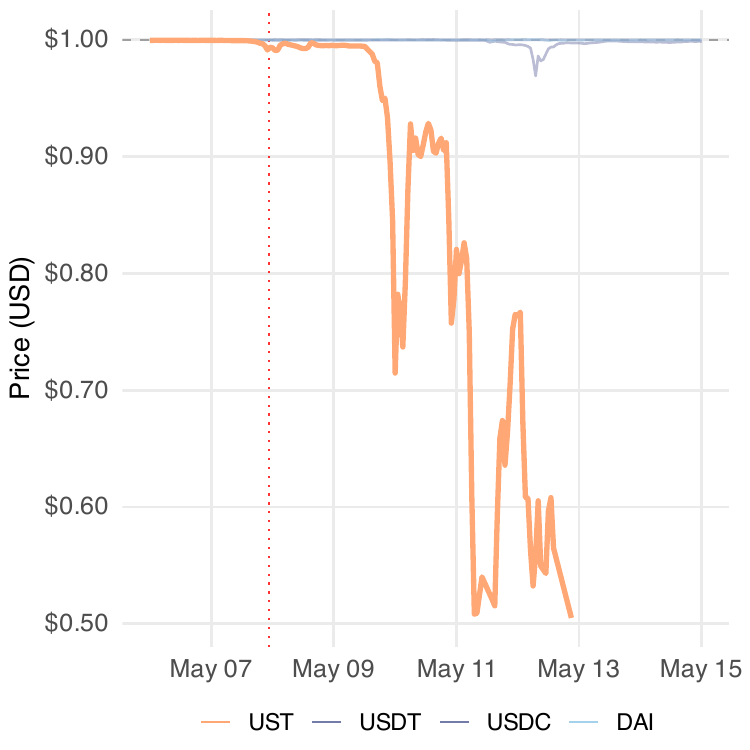} &
\includegraphics[width=0.34\textwidth]{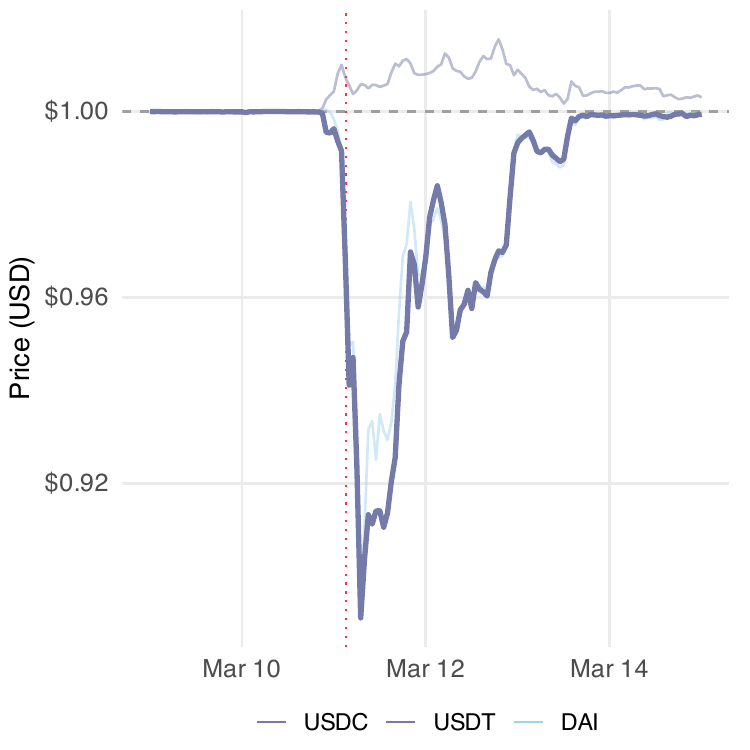} \\
(a) UST Collapse (May 2022) & (b) SVB/USDC Crisis (Mar 2023) \\[0.5em]
\includegraphics[width=0.34\textwidth]{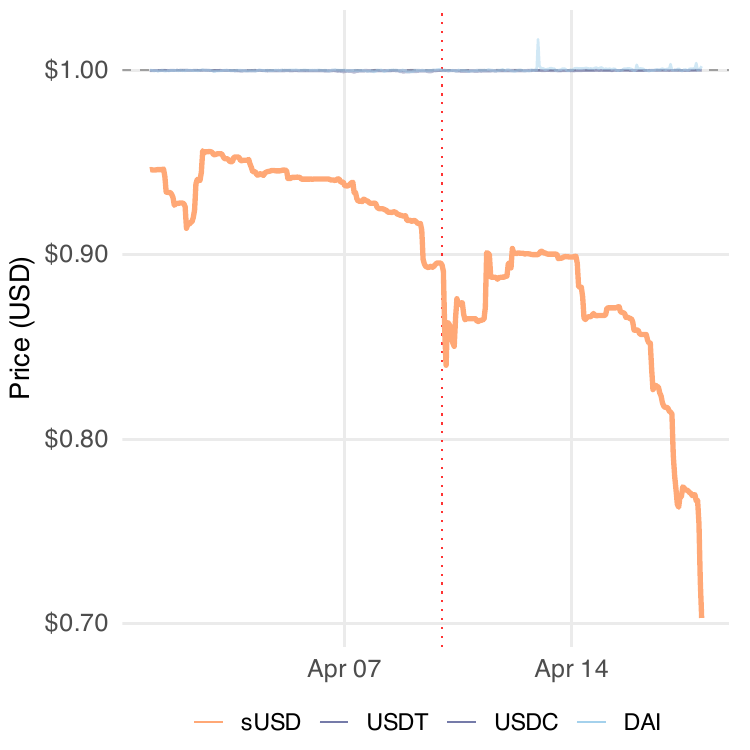} &
\includegraphics[width=0.34\textwidth]{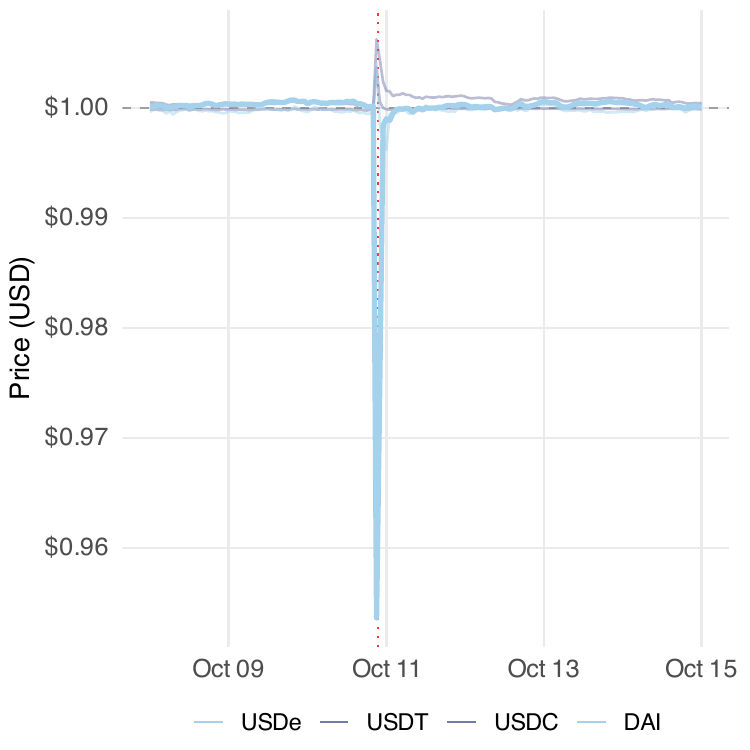} \\
(c) sUSD Depeg (Apr 2025) & (d) USDe Crash (Oct 2025)
\end{tabular}
\caption{Price paths during major depeg events. Vertical dashed lines indicate event dates. The affected stablecoin is shown with a thicker line. UST and USDe experienced severe crashes ($>$30\% depeg), while USDC and sUSD showed more moderate deviations.}\label{fig:event_prices}
\end{figure}

We analyze four events spanning all mechanism types: (i) UST Collapse (May 2022, algorithmic), where UST fell to \$0.50; (ii) SVB/USDC Crisis (March 2023, fiat-backed), where USDC dropped to \$0.87; (iii) sUSD Depeg (April 2025, algorithmic); and (iv) USDe Crash (October 2025, crypto-collateralized), where USDe fell to \$0.68 during a \$19 billion liquidation cascade. Figure~\ref{fig:event_prices} displays price paths, illustrating heterogeneous depeg dynamics.

Our identification strategy proceeds in two steps. First, we apply the Forbes-Rigobon adjustment \citep{forbes2002contagion} to distinguish true contagion from volatility-induced correlation increases:
\begin{equation}
\rho_{adj} = \frac{\rho_{crisis}}{\sqrt{1 + \delta(1 - \rho_{crisis}^2)}}
\end{equation}
where $\delta = (\sigma^2_{crisis}/\sigma^2_{calm}) - 1$. True contagion is identified when adjusted correlations significantly exceed pre-event levels. Second, we exploit the SVB crisis as a natural experiment using synthetic control methods.

Table~\ref{tab:event_spillover} documents that all four events generate substantial spillover increases, but with notable heterogeneity. The USDe crash (+53.1 pp) and UST collapse (+26.4 pp) produced the largest system-wide effects, while the sUSD depeg (+0.5 pp) generated almost none. This difference reflects market structure: sUSD's small size (\$50M) and pre-existing chronic depeg limited its systemic impact, while USDe's \$6.6B market cap and sudden collapse triggered cascading liquidations.

\begin{table}[!t]
\caption{Total spillover index changes during major depeg events\label{tab:event_spillover}}
\begin{tabular*}{\columnwidth}{@{\extracolsep\fill}lccccc@{\extracolsep\fill}}
\toprule
Event & Affected & Mechanism & Pre-Event & During & $\Delta$ \\
\midrule
UST Collapse & UST & Algorithmic & 6.4\% & 32.8\% & +26.4 \\
SVB/USDC & USDC & Fiat-backed & 7.3\% & 37.3\% & +30.1 \\
sUSD Depeg & sUSD & Algorithmic & 20.3\% & 20.7\% & +0.5 \\
USDe Crash & USDe & Crypto-coll. & 17.3\% & 70.4\% & +53.1 \\
\bottomrule
\end{tabular*}
\begin{tablenotes}
\item Note: Total spillover index at $\tau = 0.5$ (median). Pre-event is 30-day estimation window; During is event window. High-frequency (hourly) QVAR estimation. sUSD's minimal increase reflects its small market cap and pre-existing chronic depeg.
\end{tablenotes}
\end{table}

The Forbes-Rigobon tests reveal the mechanism-specific heterogeneity that is our central finding. While raw correlations spike during all events, the \textit{residual} contagion after volatility adjustment varies systematically by mechanism type.

\begin{figure}[!t]
\centering
\begin{tabular}{cc}
\includegraphics[width=0.34\textwidth]{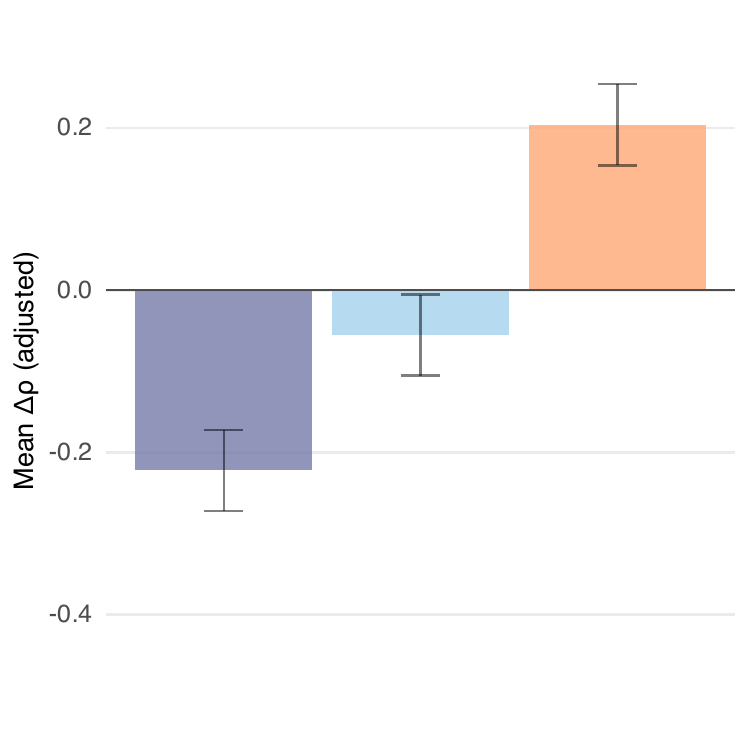} &
\includegraphics[width=0.34\textwidth]{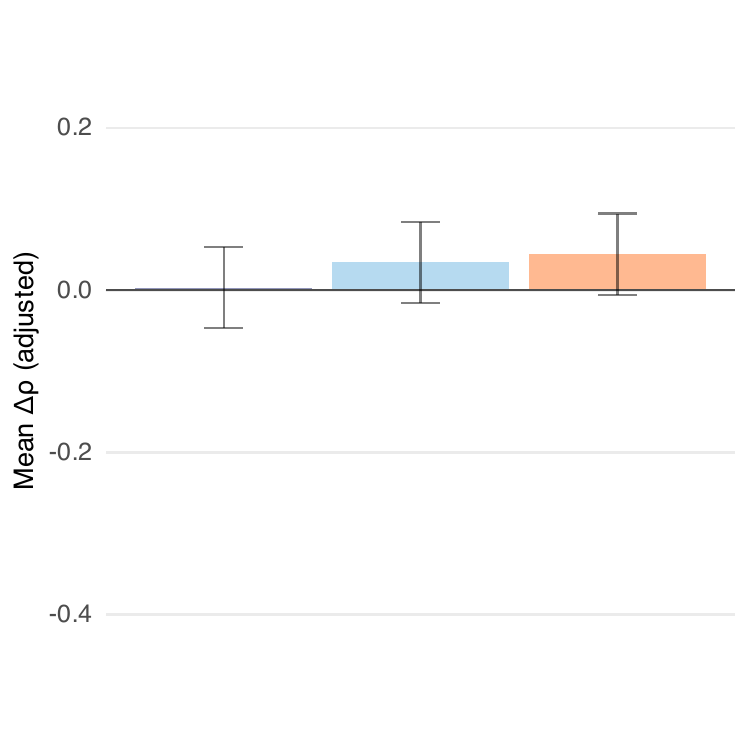} \\
(a) UST Collapse (May 2022) & (b) SVB/USDC Crisis (Mar 2023) \\[0.5em]
\includegraphics[width=0.34\textwidth]{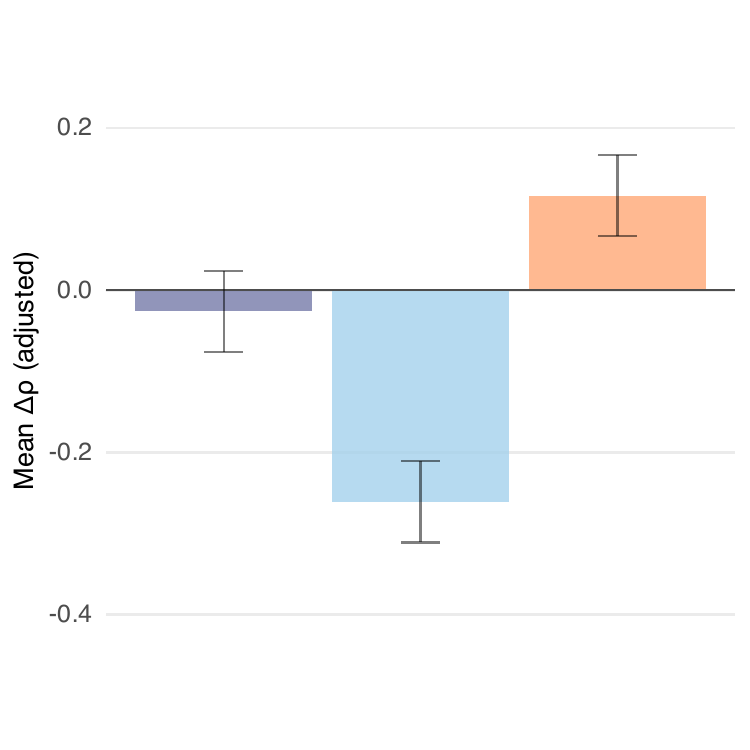} &
\includegraphics[width=0.34\textwidth]{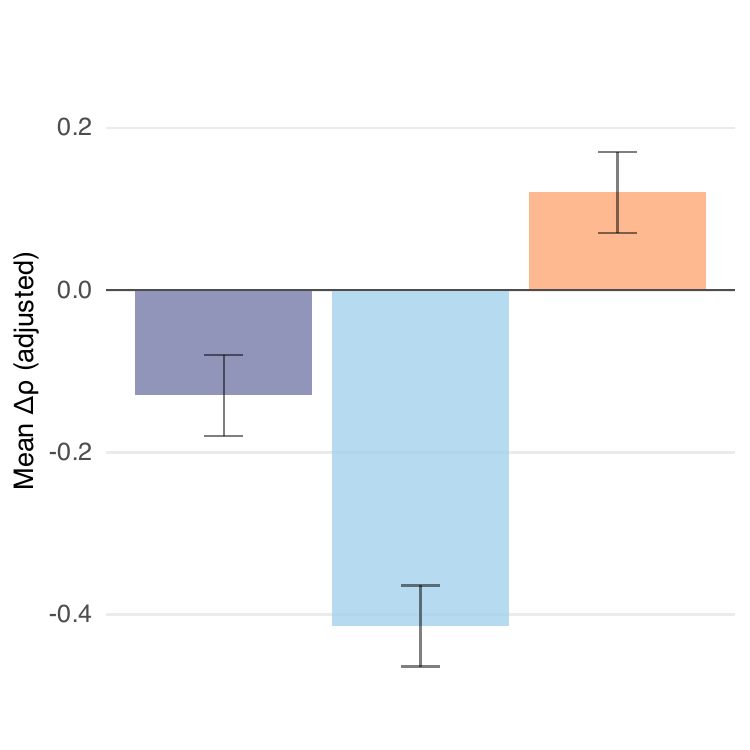} \\
(c) sUSD Depeg (Apr 2025) & (d) USDe Crash (Oct 2025)
\end{tabular}
\caption{Forbes-Rigobon adjusted correlation changes by stablecoin category during depeg events. Bars represent mean $\Delta\rho_{adj}$ from affected coin to target category. Colors: Fiat-backed (purple), Crypto-collateralized (blue), Algorithmic (orange).}\label{fig:event_contagion}
\end{figure}

Figure~\ref{fig:event_contagion} displays adjusted correlation changes by target category. The pattern reveals mechanism-specific heterogeneity: algorithmic stablecoins exhibit positive adjusted changes (mean $\Delta\rho_{adj} = +0.07$), indicating genuine contagion beyond volatility effects. In contrast, both fiat-backed (mean $\Delta\rho_{adj} = -0.10$) and crypto-collateralized stablecoins (mean $\Delta\rho_{adj} = -0.14$) show negative adjusted changes, reflecting flight-to-quality dynamics where investors rotate toward perceived safe havens during stress. This heterogeneity confirms that stabilization design shapes crisis response: algorithmic mechanisms amplify co-movement, while collateralized designs (whether fiat or crypto) provide relative insulation.

The SVB crisis provides a natural experiment for causal identification. Unlike the other three events, where no unexposed peers existed within the same mechanism type to serve as valid donors, the SVB shock uniquely affected USDC: Circle disclosed \$3.3 billion of reserves frozen at SVB, while other fiat-backed stablecoins had no direct exposure. We exploit this differential using synthetic control methods \citep{abadie2010synthetic}, constructing a counterfactual ``synthetic USDC'' from unexposed coins (BUSD: 52.1\%, USDT: 27.7\%, TUSD: 20.3\%) matched to USDC's pre-event spillover trajectory.

\begin{figure}[!t]
\centering
\includegraphics[width=0.95\textwidth]{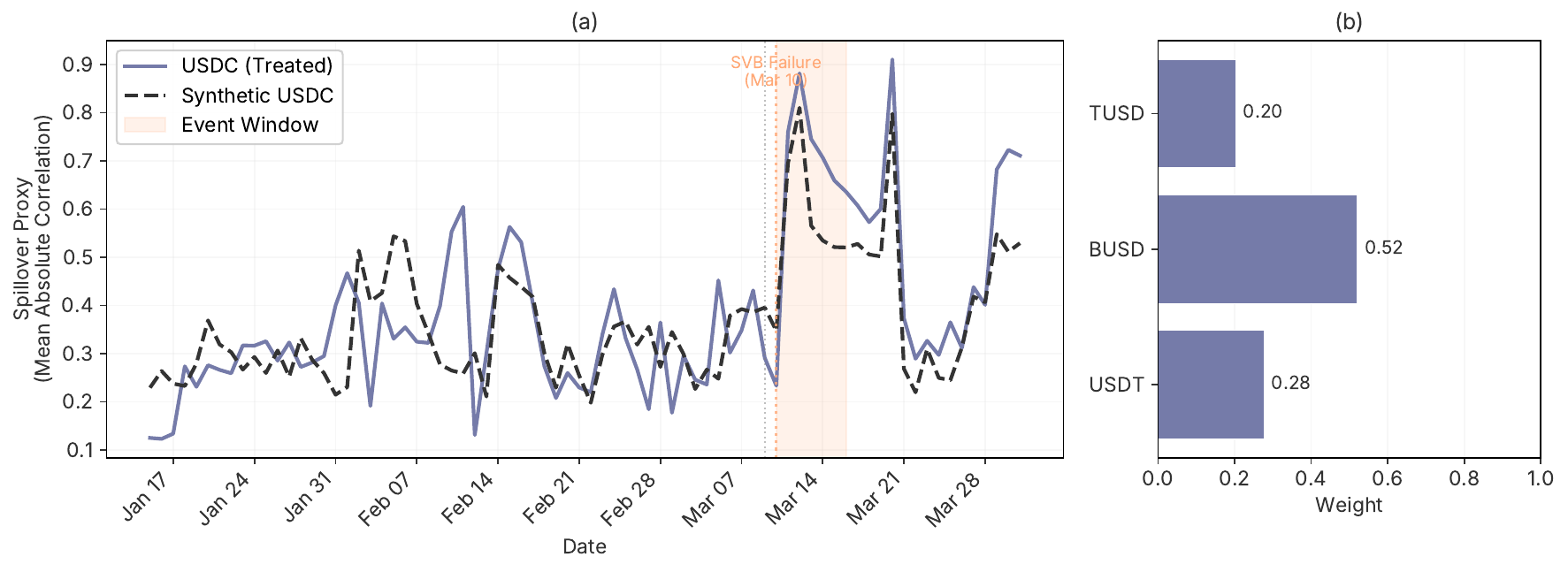}
\caption{Synthetic control analysis of SVB/USDC crisis. (a) Spillover proxy for USDC (treated) versus synthetic USDC; shaded region indicates event window. (b) Synthetic control weights. Pre-event RMSE = 0.116; treatment effect = +0.09 (p = 0.75, placebo inference).}\label{fig:svb_synth}
\end{figure}

Figure~\ref{fig:svb_synth} displays the results. The pre-event fit is RMSE = 0.116, validating the synthetic control construction. During the crisis, both USDC and its synthetic counterpart experience sharp spillover increases, but USDC's excess effect is small (+0.09) and statistically insignificant. Placebo tests yield a rank-based p-value of 0.75: therefore USDC's effect is indistinguishable from noise.

Therefore, the SVB shock did not transform USDC into a systemic risk amplifier. While USDC's price crashed to \$0.87, its \textit{contagion behavior} was indistinguishable from unexposed peers. This provides causal evidence that fiat-backed stablecoins function as stability anchors even when directly exposed to banking sector stress.

We acknowledge that the three-coin donor pool limits statistical power. However, the null result is itself informative: even with limited precision, we can reject large treatment effects. The 95\% confidence interval for USDC's excess spillover is $[-0.15, +0.33]$, ruling out the hypothesis that SVB exposure transformed USDC into a major risk amplifier.

\subsection{Robustness Analysis}\label{sec:robustness}

Table~\ref{tab:robustness} reports sensitivity analyses across lag orders of 1, 2, and 3, forecast horizons of 5, 10, 15, and 20 days, alternative quantile definitions, and different sample periods. The core finding, that tail spillovers exceed median conditions, is robust across all specifications. Median spillovers remain stable at 12.8--14.1\%, while tail spillovers vary with parameter choices but consistently exceed median levels by 10--50 percentage points.

\input{tables/robustness_table.tex}

Subsample analysis addresses whether results are driven by the UST collapse. We split the sample into a pre-UST period from April 2021 to April 2022 with 369 observations, and a post-UST period from June 2022 to December 2025 with 1,243 observations. Both show elevated tail spillovers relative to median conditions: left-tail spillovers of 28--31\% versus median spillovers of 15--16\%. This pattern persists across both subsamples, confirming that mechanism-specific heterogeneity is not an artifact of any single crisis episode.

\subsection{Limitations}\label{sec:limitations}

Three limitations warrant acknowledgment. First, data constraints exclude UST, FDUSD, and USDe from the main QVAR analysis; the event studies partially address this by examining these coins using high-frequency data. Second, while we strengthen causal identification through Forbes-Rigobon adjustments and synthetic control methods, the small donor pool of three doners for the SVB analysis limits statistical power. Third, our sample period captures a specific market regime; spillover patterns may differ under alternative regulatory frameworks or market structures.

\section{Conclusions and Policy Implications}\label{sec:conclusion}

\subsection{Conclusions}

This paper provides quantile-based evidence on how stablecoin mechanism design shapes systemic risk behavior. Three findings emerge from our analysis.

First, QVAR estimates at the 5th, 50th, and 95th percentiles reveal that fiat-backed stablecoins maintain near-zero net spillovers across all quantiles, while algorithmic and crypto-collateralized designs show 15--50 percentage point amplification specifically in the tails. This magnitude is invisible to median-based analysis.

Second, our three-tier fiat-stablecoin-crypto network analysis reveals that the theoretical isolation between traditional and crypto markets breaks down under stress. Direct volatility channels between the US Dollar Index and Bitcoin activate during tail events, bypassing stablecoin intermediation entirely.

Third, Forbes-Rigobon adjusted correlations correct for heteroskedasticity bias that inflates crisis-period correlations. After this adjustment, algorithmic stablecoins exhibit genuine contagion with a mean adjusted correlation change of +0.07, while collateralized designs show flight-to-quality effects: fiat-backed coins average $-$0.10 and crypto-collateralized coins average $-$0.14. A synthetic control analysis of the SVB crisis confirms that fiat-backed stablecoin spillovers are statistically indistinguishable from unexposed peers at p = 0.75, even when directly exposed to banking sector stress.

These findings imply that uniform stablecoin regulation is inappropriate. Mechanism-specific frameworks, including differentiated capital requirements, stress testing protocols, and disclosure standards, are necessary to address the heterogeneous systemic risk profiles documented here.

\subsection{Policy Implications}

Our findings support mechanism-specific regulatory frameworks. The 15--50 percentage point increase in tail spillovers relative to normal conditions (Table~\ref{tab:robustness}) implies that static risk measures substantially underestimate systemic exposure during stress. Three specific recommendations follow:

First, \textit{differentiated capital requirements} by mechanism type. Fiat-backed stablecoins' near-zero net spillovers justify bank-like regulation focused on reserve adequacy. Algorithmic and crypto-collateralized designs require additional buffers; our estimates suggest regulatory capital buffers for extreme losses should be 2--3$\times$ higher than median-based measures would indicate.

Second, \textit{dynamic margin requirements} for exchanges. The breakdown of fiat-crypto isolation under stress (Figure~\ref{fig:extended_spillover_network}) means that margin models assuming independent asset classes will fail precisely when they matter most. Platforms should implement quantile-conditional margins that increase automatically when spillover indices exceed historical norms.

Third, \textit{mechanism-specific stress testing}. Our Forbes-Rigobon results show that algorithmic coins exhibit contagion while collateralized designs show flight-to-quality, implying that uniform stress scenarios are inappropriate. Regulators should require issuers to model failure modes specific to their mechanism: death spirals for algorithmic designs, banking sector contagion for fiat-backed reserves, and collateral liquidation cascades for crypto-collateralized designs.

\section*{Funding and Conflict of Interest}

This research received no specific grant from any funding agency in the public, commercial, or not-for-profit sectors. The authors declare no conflict of interest.

\section*{Data Availability}

Price data are sourced from CoinMetrics Pro, a proprietary subscription service; due to license restrictions, the raw data cannot be publicly shared but are available to authorized users through the CoinMetrics platform. The US Dollar Index (DTWEXB) is publicly available from the Federal Reserve Economic Data (FRED) database.
Replication code for the QVAR analysis and event studies is available at \url{https://github.com/dthinkr/stablecoin-spillover} upon publication.

\bibliographystyle{abbrvnat}

\input{main.bbl}
\end{document}

%% file: tables/spillover_methods_table.tex
\begin{table}[!t]
\centering
\caption{Comparison of research methods for stablecoin spillover effects}
\label{tab:spillover_methods}
\footnotesize
\begin{tabular*}{\columnwidth}{@{\extracolsep\fill}p{2.2cm}p{2.8cm}p{2.8cm}p{2.2cm}p{3.5cm}@{\extracolsep\fill}}
\toprule
\textbf{Study} & \textbf{Methodology} & \textbf{Market Scope} & \textbf{Spillover Metrics} & \textbf{Key Findings} \\
\midrule
\citet{fernandez-mejia2024ig} & Cross-Quantilogram & Stablecoins, financial \& crypto indices & Price volatility, extreme changes & Asymmetry in stablecoin price responses \\
\addlinespace
\citet{giudici2020ke} & Volatility spillover, VAR & Foreign currencies, stablecoins & Volatility spillover, currency shocks & Basket-based stablecoins less volatile \\
\addlinespace
\citet{gubareva2023xn} & Quantile causality, TVP-VAR & Stocks, bonds, stablecoins, crypto & Trading volume, market cap & ``Crypto flight to safety'' phenomenon \\
\addlinespace
\citet{huang2023jk} & Copula-CoVaR, ARMA-GARCH & Stablecoins, crypto, non-crypto & Risk spillover, CoVaR & Significant cross-market risk spillovers \\
\addlinespace
\citet{iyer2023new} & Barun\'ik-K\v{r}ehl\'ik (BK) & Crypto assets, stock markets & Price volatility, returns & Enhanced crypto-stock interconnectedness \\
\addlinespace
\citet{jausyan2023mr} & GARCH, TVP-VAR & Stablecoins, unbacked crypto, financial & Systemic impact, volatility & Stablecoins have significant systemic impact \\
\addlinespace
\citet{kim2022dk} & Two-stage least squares & Stablecoins, commercial paper, Treasury & Issuance, yields & Issuance affects CP and Treasury yields \\
\addlinespace
\citet{kolodziejczyk2023hc} & Quantile coherence & Stablecoins, crypto, stock indices & Hedging, safe-haven & Stablecoins as weak hedges, safe havens \\
\addlinespace
\citet{paeng2024xk} & Quantile Granger causality & S\&P 500, stablecoins, crypto & Causality, spillovers & Bidirectional causality and spillovers \\
\addlinespace
\citet{let2023gy} & Spectral variance decomp. & Volatile crypto, stablecoins & Spillovers, investor activity & Crypto shocks drive stablecoin popularity \\
\botrule
\end{tabular*}
\end{table}

%% file: tables/depeg_events_table.tex
\begin{table}[htbp]
\caption{Major stablecoin depeg events and their causes\label{tab:depeg_events}}
\begin{tabular*}{\columnwidth}{@{\extracolsep\fill}llrll@{\extracolsep\fill}}
\toprule
Stablecoin & Date & Deviation & Type & Cause \\
\midrule
\multicolumn{5}{l}{\textit{Fiat-backed}} \\
USDC & Mar 2023 & $-$5.9\% & Macro & SVB collapse; \$3.3B reserves frozen \\
DAI & Mar 2023 & $-$4.8\% & Contagion & Spillover from USDC via PSM \\
\midrule
\multicolumn{5}{l}{\textit{Crypto-collateralized}} \\
LUSD & Mar 2022 & $-$23.3\% & Macro & Fed tightening; ETH collateral decline \\
USDe & Oct 2025 & $-$32.0\% & Macro & \$19B liquidation cascade; oracle failure \\
\midrule
\multicolumn{5}{l}{\textit{Algorithmic}} \\
UST & May 2022 & $-$37.7\% & Endogenous & Death spiral; Luna hyperinflation \\
FRAX & Jun 2022 & +15.2\% & Contagion & Post-Terra flight to quality \\
sUSD & Apr 2025 & $-$22.6\% & Endogenous & SIP-420 removed arbitrage incentives \\
\botrule
\end{tabular*}
\begin{tablenotes}
\item Note: Events identified from deviation spikes in our dataset; causes attributed through contemporaneous news sources and protocol documentation. Deviations represent maximum price deviation from \$1 peg. Events are annotated in Figure~\ref{fig:stablecoin_diff}.
\end{tablenotes}
\end{table}

%% file: tables/robustness_table.tex
\begin{table}[!t]
\caption{Robustness of total spillover index to parameter choices\label{tab:robustness}}
\begin{tabular*}{\columnwidth}{@{\extracolsep\fill}lccc@{\extracolsep\fill}}
\toprule
Specification & $\tau = 0.05$ & $\tau = 0.50$ & $\tau = 0.95$ \\
\midrule
\multicolumn{4}{l}{\textit{Panel A: Lag Order (H = 10)}} \\
$p = 1$ (baseline) & 28.9 & 13.8 & 24.6 \\
$p = 2$ & 35.7 & 12.8 & 38.5 \\
$p = 3$ & 44.8 & 13.4 & 35.4 \\
\midrule
\multicolumn{4}{l}{\textit{Panel B: Forecast Horizon (p = 1)}} \\
$H = 5$ & 18.1 & 13.1 & 16.8 \\
$H = 10$ (baseline) & 28.9 & 13.8 & 24.6 \\
$H = 15$ & 39.9 & 14.0 & 30.1 \\
$H = 20$ & 51.0 & 14.1 & 34.1 \\
\midrule
\multicolumn{4}{l}{\textit{Panel C: Quantile Choice (p = 1, H = 10)}} \\
$\tau \in \{0.01/0.5/0.99\}$ & 66.5 & 13.8 & 51.6 \\
$\tau \in \{0.05/0.5/0.95\}$ (baseline) & 28.9 & 13.8 & 24.6 \\
$\tau \in \{0.1/0.5/0.9\}$ & 20.5 & 13.8 & 16.2 \\
\botrule
\end{tabular*}
\begin{tablenotes}
\item Note: Total spillover index (\%) under different parameter specifications. Baseline specification: lag order $p=1$, forecast horizon $H=10$, quantiles $\tau \in \{0.05, 0.50, 0.95\}$. Results based on 1,643 daily observations (April 2021--December 2025) for 10 assets (8 stablecoins, Bitcoin, US Dollar Index).
\end{tablenotes}
\end{table}